\documentclass[prb,aps,groupedaddress,amsmath,twocolumn]{revtex4-1}
\usepackage[dvips]{graphicx}
\usepackage{dcolumn}
\usepackage{bm}
\usepackage{color}

\begin{document}

\title{Fate of density functional theory in high-pressure solid hydrogen}

\author{Sam Azadi}
\affiliation{Thomas Young Centre and Department of Physics, Imperial College London, 
London SW7 2AZ, United Kingdom}
\author{W. M. C. Foulkes}
\affiliation{Thomas Young Centre and Department of Physics, Imperial College London,
London SW7 2AZ, United Kingdom}


\begin{abstract}
  This paper investigates some of the successes and failures of density
  functional theory in the study of high-pressure solid hydrogen at low
  temperature. We calculate the phase diagram, metallization pressure,
  phonon spectrum, and proton zero-point energy using three popular
  exchange-correlation functionals: the local density approximation
  (LDA), the Perdew-Burke-Ernzerhof (PBE) generalized gradient
  approximation, and the semi-local Becke-Lee-Yang-Parr (BLYP)
  functional. We focus on the solid molecular P$6_3$/m, C2/c, Cmca-12,
  and Cmca structures in the pressure range from $100<P<500$ GPa over
  which phases I, II and III are observed experimentally.  At the static
  level of theory, in which proton zero-point energy is ignored, the
  LDA, PBE and BLYP functionals give very different structural
  transition and metallization pressures, with the BLYP phase diagram in
  better agreement with experiment.  Nevertheless, all three functionals
  provide qualitatively the same information about the band gaps of the
  four structures and the phase transitions between them. Going beyond
  the static level, we find that the frequencies of the vibron modes
  observed above 3000~cm$^{-1}$ depend strongly on the choice of
  exchange-correlation functional, although the low-frequency part of
  the phonon spectrum is little affected.  The largest and smallest
  values of the proton zero-point energy, obtained using the BLYP and
  LDA functionals, respectively, differ by more than 10 meV/proton.
  Including the proton zero-point energy calculated from the phonon
  spectrum within the harmonic approximation improves the agreement of
  the BLYP and PBE phase diagrams with experiment.  Taken as a whole,
  our results demonstrate the inadequacy of mean-field-like density
  functional calculations of solid molecular hydrogen in phases I, II
  and III and emphasize the need for more sophisticated methods.
\end{abstract}

\maketitle

\section {Introduction}

In 1935, Wigner and Huntington \cite{wigner} predicted that the
molecules in solid molecular hydrogen at very high pressure would
dissociate to form a metallic atomic solid. Solid hydrogen is expected
to transform into a metallic liquid ground state at high
pressure,\cite{Bonev} may exhibit high-T$_\text{c}$ superconductivity
and superfluidity,\cite{Ashcroft1, Ashcroft2, Richardson} and plays an
important role in astrophysics.\cite{Alavi} Although it is not yet
possible to reach the static pressure of more than 400 GPa normally
thought necessary to dissociate the hydrogen molecules, recent
experimental results obtained using diamond anvil cell techniques have
been interpreted as indicating metallization and the transition of
molecular hydrogen to an atomic liquid, known as phase IV, below 300
GPa.\cite{Eremets} Other experimental groups disagree, observing no
evidence of the optical conductivity expected of a metal over the entire
range of temperatures to the highest pressures explored, and no
signature of an atomic liquid.\cite{Hemley12} The interpretation of
experimental results is complicated by the fact that, unlike phase I
($<110$ GPa), which is a molecular solid of quantum rotors on a
hexagonal close packed lattice, the structures of phase II (known as the
broken-symmetry phase\cite{Kohanoff}) and phase III ($>150$ GPa) are
unknown.\cite{Akahama} In addition, it remains unclear whether or not
phase III is metallic.\cite{Hemley96}

Metallization is believed to occur either via the dissociation of
hydrogen molecules and a structural transformation to an atomic metallic
phase or via band overlap (band-gap closure) within the molecular
phase. There is more evidence in favor of the latter assumption.
Quantum Monte Carlo (QMC) calculations of metallic hydrogen
\cite{Natoli} and our knowledge of the experimental equation of
state\cite{Loubeyre1} suggest that the hydrogen molecules remain
undissociated at pressures up to 620 GPa. Density functional theory
(DFT) calculations using the ab initio random structure searching
approach\cite{PickardReview} predict dissociation near 500
GPa,\cite{Pickard} whilst DFT calculations using evolutionary
techniques\cite{oganov1} predict that the formation of stable
non-molecular metallic hydrogen requires pressures much higher than 600
GPa.\cite{oganov3} These values are so large that it seems likely that
metallization takes place by band-gap closure in the molecular phase
well before dissociation occurs.

Experimental vibrational and scattering results are not yet sufficient
to allow an unambiguous determination of the structure of phase III,
leaving theorists free to speculate.\cite{Alder, Chacham, Kaxiras,
  Tosatti, Johnson, Martin} However, the range of structures proposed is
so wide and the energy differences between them so small (typically a
few meV/proton) that it is unclear which, if any, corresponds to the
global free energy minimum. Finding the most stable arrangements of
atoms in solids and molecules is and will remain a very difficult task,
but significant progress has been made over the past few years. Recent
advances based on simulated annealing, metadynamics, random sampling,
evolutionary algorithms, basin hopping, minima hopping, and data mining,
are discussed in Ref.~\onlinecite{oganov2}.

The ab initio random structure searching method\cite{PickardReview} has
also been used to investigate the zero-temperature phase diagram of
molecular solid hydrogen.\cite{Pickard} Searches excluding the effects
of proton zero-point energy (ZPE) found a sequence of molecular crystal
structures within the pressure range over which phase III is observed
experimentally. The C2/c structure was stable up to 270 GPa; Cmca-12 was
stable from 270 GPa to 385 GPa; and Cmca was stable above 385 GPa. The
consequences of including proton ZPE (see Section~IV) were investigated
only after the structures had been found. Ref.~\onlinecite{Pickard}
provided many new ideas about the phase diagram of solid molecular
hydrogen, but the energies of the crystal structures considered were
obtained using a simple semi-local approximation to the unknown
exchange-correlation (XC) functional of density functional theory (DFT),
which may or may not be accurate. Moreover, since DFT substantially
underestimates the fundamental band gap for essentially all materials,
it also underestimates the pressure at which the band gap
closes.\cite{Mujica} Finally, it is worth noting that crystal structures
obtained using random searching do not always agree with structures
obtained using other techniques or experiment.\cite{Martinez, Ma} This
is not surprising, as the probability that a random search will find the
ground state decreases exponentially with increasing system
size.\cite{oganov2}

DFT is one of very few quantum mechanical methods capable of calculating
the energies and enthalpies of all phases of interest over all relevant
pressure regimes and has played an important role in studies of
high-pressure hydrogen. However, serious doubts about the accuracy of
the results persist. How do the inter-atomic interaction energy, the
bond-stretch energy, the phase diagram, the metallization mechanism, and
the phonon spectrum depend on the approximation used for the XC
functional?  How accurate should we expect DFT calculations of measured
quantities such as infra-red (IR) and Raman spectra to be? Answering
these questions is necessary to assess the reliability of the many
existing DFT simulations of high-pressure solid hydrogen.\cite{me1, me2}
On the minus side, because the enthalpy differences between rival
structures are so small (a few meV/proton), it is very difficult to be
sure that a specific structure represents the global minimum.  The
failures of DFT to describe water and other systems containing hydrogen
at low pressure and temperature have already been
documented.\cite{gillan} Perhaps surprisingly, the errors of DFT are
already troublesome even for a single H$_2$O molecule: it was shown
recently \cite{santra, cohen} that some common XC functionals provide
rather poor estimates of the O--H bond-stretch energy. Because of the
limited experimental data, theoretical and computational studies have an
unusually important role to play in the study of solid hydrogen, but
only if they are correct.

Our purpose in this paper is to investigate the successes and failures
of several widely used XC functionals in the study of high-pressure
solid hydrogen.  We emphasize that the failures we identify do not
signal a breakdown of DFT itself; they demonstrate only the limited
accuracy of certain XC functionals. We believe that identifying and
investigating deficiencies such as these provides useful information to
developers of new XC functionals and may bring us closer to one of the
main goals of DFT, which is to design functionals capable of providing
accurate total energies for as wide a variety of systems as possible.

Starting at the static level of theory, we calculate the
zero-temperature phase diagram using three different functionals: the
Perdew-Zunger (PZ) parameterization of the local density
approximation\cite{LDA} (LDA), the Perdew-Burke-Ernzerhof (PBE) version
of the generalized gradient approximation\cite{PBE} (GGA), and the
semi-local Becke-Lee-Yang-Parr (BLYP) functional.\cite{BLYP} We focus on
four specific structures with space groups P$6_3$/m, C2/c, Cmca-12, and
Cmca. According to previous static DFT calculations using the PBE
functional,\cite{Pickard} these are stable in the pressure ranges $<$
105 GPa, 105--270 GPa, 270--385 GPa, and 385--490 GPa, respectively.

Nuclear quantum effects are generally neglected in DFT calculations
because of computational cost. However, there is strong evidence to show
that this neglect has a significant impact on the results for solid
hydrogen, even at finite temperature.\cite{Morales} We investigate
quantum nuclear effects at zero temperature by calculating the proton
ZPE as a function of pressure within the quasi-harmonic approximation.
Our calculations of the phonon spectrum also enable us to investigate
the effect of the choice of XC functional on the frequencies of the
Raman and IR active modes. We demonstrate that adding the proton ZPE to
the calculated enthalpy produces ``dynamic'' phase diagrams in better
agreement with experiment than the static phase diagrams obtained
excluding ZPE.

The paper is organized as follows. Section II describes the details of
our DFT calculations.  Section III investigates the effect of
the choice of XC functional on the calculated phase diagram and
metallization pressure of solid molecular hydrogen at the static level
of theory, and shows how the choice of functional affects the pressures
at which the phase transitions I $\rightarrow$ II $\rightarrow$ III are
predicted to take place.  In Section IV, we go beyond the static level
by investigating the effects of ZPE.  Section VI concludes.

\section {Computational Details}

Since the energy differences between high-pressure solid molecular
structures are very small, the calculations must be done with the
highest possible numerical precision.  Our DFT calculations were carried
out within the pseudopotential and plane-wave approach using the Quantum
Espresso suite of programs.\cite{QS} All calculations used
norm-conserving pseudopotentials and a basis set of plane waves with a
cutoff of 200 Ry.  Geometry and cell optimizations employed a dense
$16\times16\times16$ ${\bf k}$-point mesh. The BFGS quasi-Newton
algorithm was used for cell and geometry optimization, with convergence
thresholds on the total energy and forces of 0.01 mRy and 0.1 mRy/Bohr,
respectively, to guarantee convergence of the total energy to better
than 1 meV/proton and the pressure to better than 0.1 GPa/proton.

To include the effects of ZPE and investigate the phonon spectrum,
vibrational frequencies were calculated using density-functional
perturbation theory as implemented in Quantum Espresso.\cite{QS} The ZPE
per proton at a specific cell volume $V$ was estimated within the
harmonic approximation: $E_{\text{ZPE}}(V) = 3 \hbar
\overline{\omega}/2$, where $\overline{\omega} = \sum_{\bf q}
\sum_{i=1}^{N_{\text{mode}}} \omega_{i}({\bf q})/(N_{\bf q}
N_{\text{mode}})$.  Here $N_{\text{mode}}$ and $N_{\bf q}$ are the
numbers of vibrational modes in the unit cell and phonon wave vectors
${\bf q}$, respectively, and the summation over ${\bf q}$ includes all
${\bf k}$-points on a $2 \times 2 \times 2$ grid in the Brillouin
zone. McMahon \cite{McMahon1, McMahon2} demonstrated that a ${\bf
  q}$-point grid of this density is sufficient to converge ZPE
differences between structures to within a few percent.  The enthalpy at
pressure $P$, including ZPE effects, was obtained by minimizing $H(P) =
E(V) + E_{\text{ZPE}}(V) + PV$ with respect to $V$ at constant $P$,
where $E(V)$ is the static ground-state energy at volume $V$.

\section {Static results and discussion}

\begin{figure}[ht]
    \centering
    \includegraphics[width=0.45\textwidth]{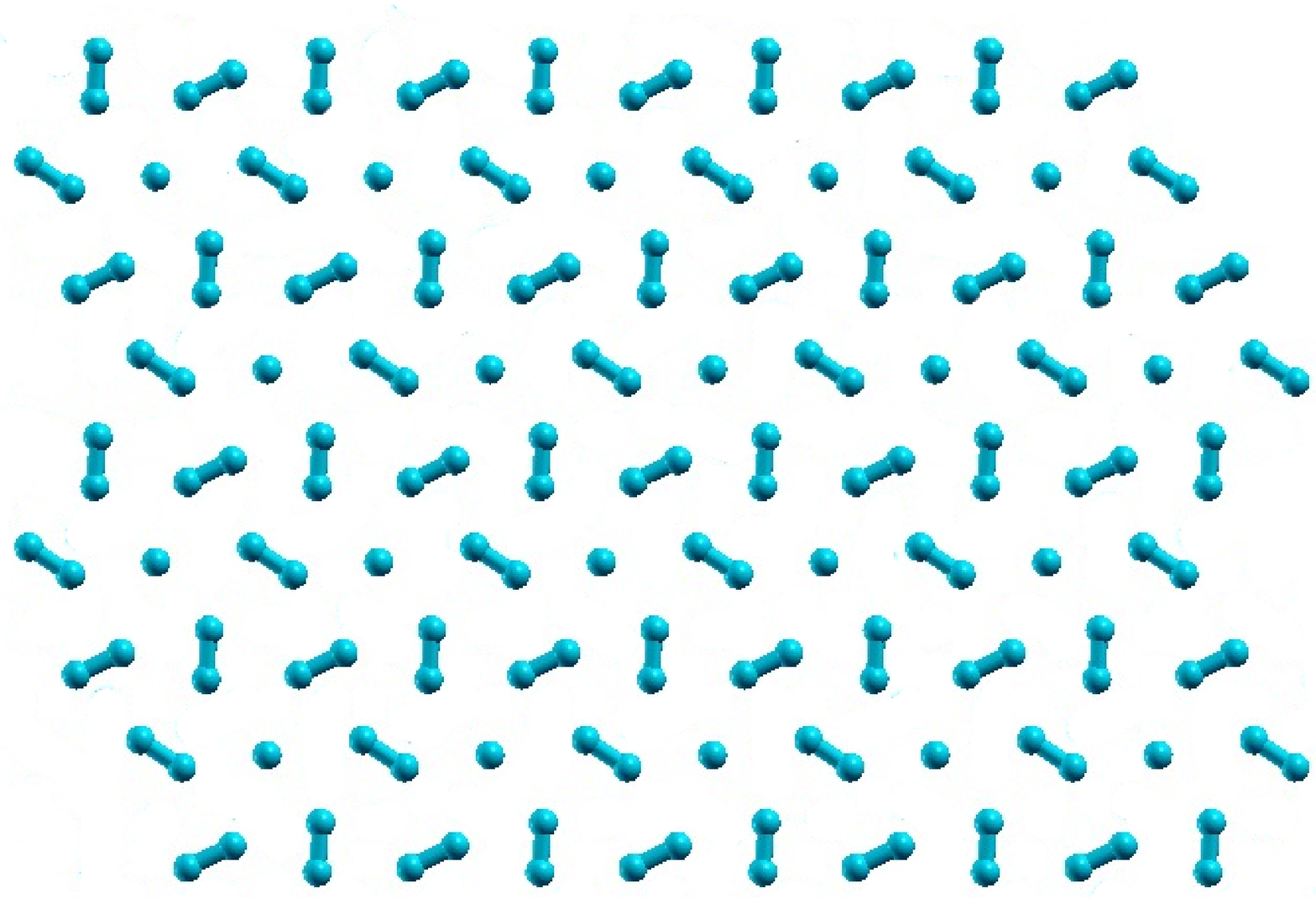} \\
    \caption{\label{P63m300}(color online) A layer of the hexagonal
      P$6_3$/m structure at 300 GPa. The layers are stacked in an ABAB
      fashion. The primitive unit cell contains 16 atoms, which form
      hydrogen molecules of two types: 75\% of the molecules lie flat
      within the plane and 25\% lie perpendicular to the plane. }
\end{figure}

\begin{figure}[ht]
    \centering
    \includegraphics[width=0.45\textwidth]{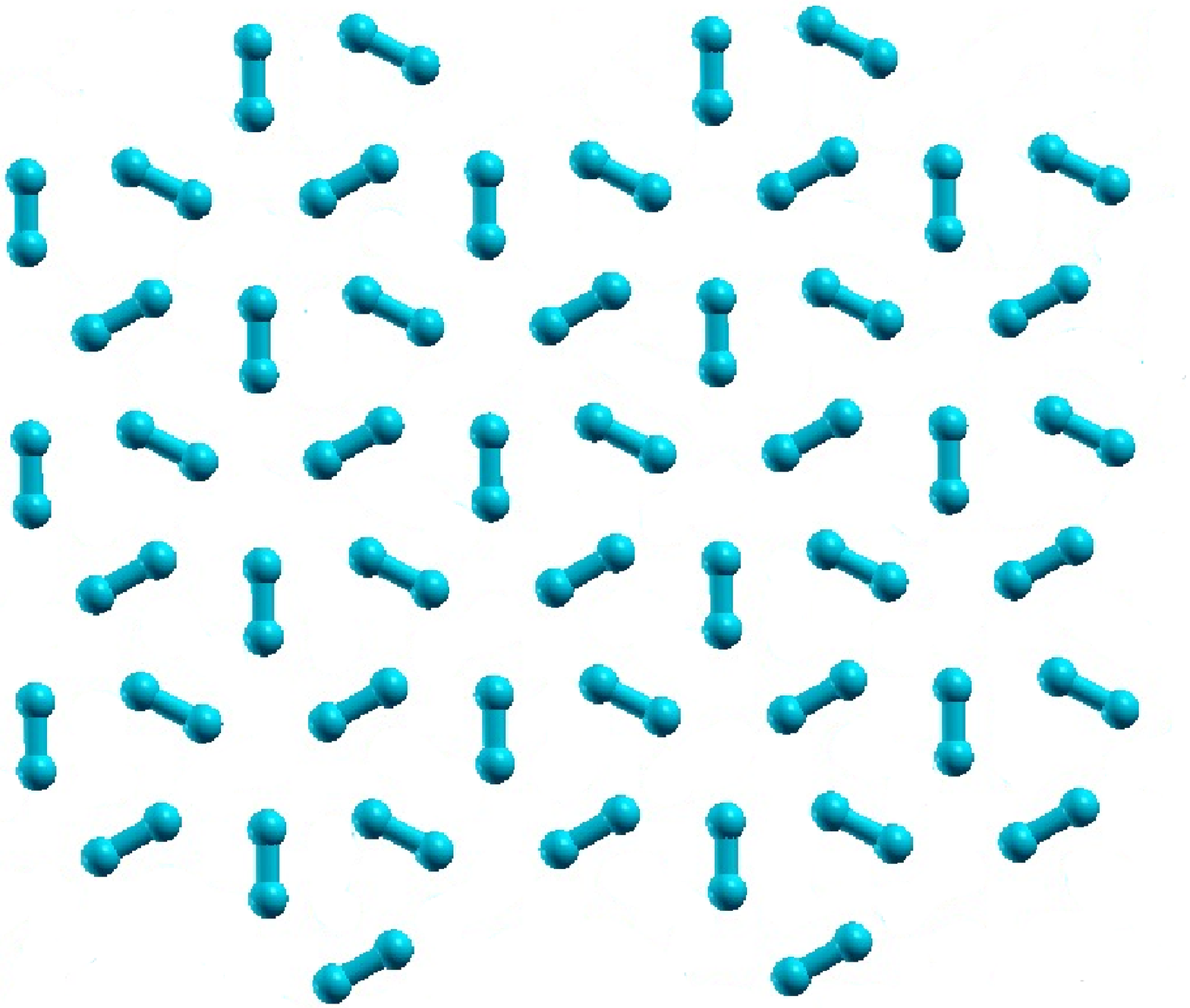} \\
    \caption{\label{C2c300}(color online) A layer of the monoclinic C2/c
      structure at 300 GPa. The layers are arranged in an ABCDA fashion
      and the primitive unit cell contains 12 atoms. }
\end{figure}

\begin{figure}[ht]
    \centering
    \includegraphics[width=0.45\textwidth]{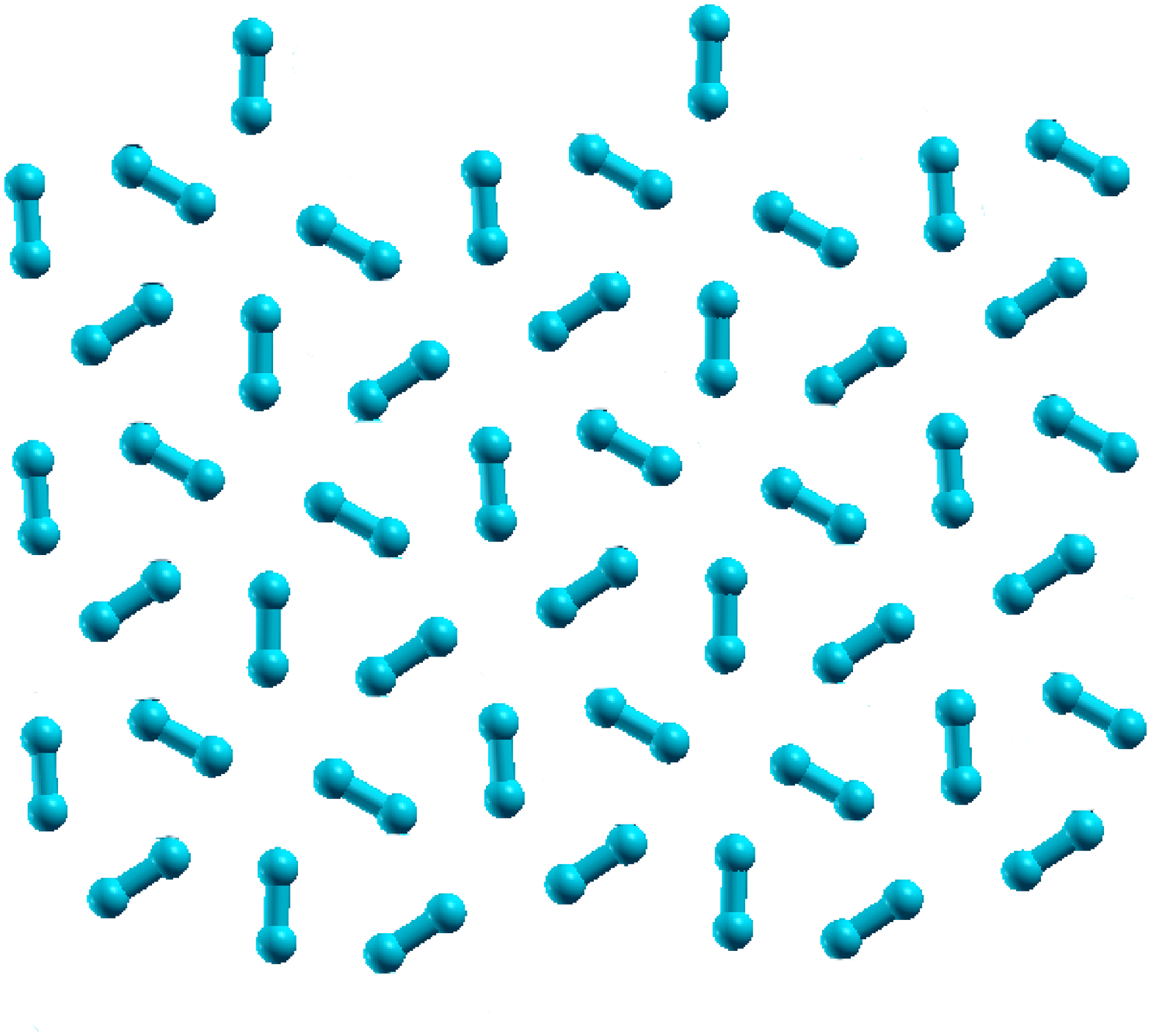} \\
    \caption{\label{Cmca12300}(color online) A layer of the monoclinic
      Cmca-12 structure at 300 GPa. The layers are arranged in an ABA
      fashion and the primitive unit cell contains 12 atoms. }
\end{figure}

\begin{figure}[ht]
    \centering
    \includegraphics[width=0.40\textwidth]{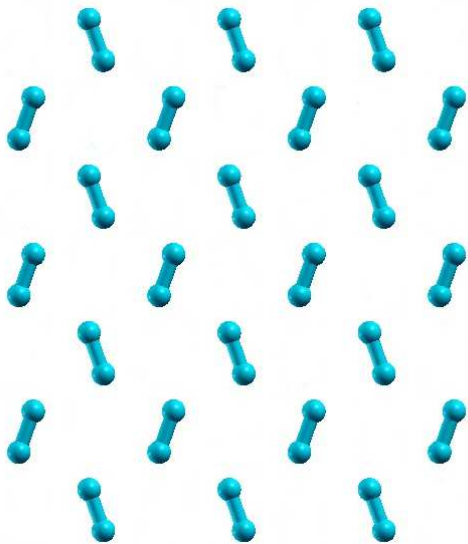} \\
    \caption{\label{Cmca300}(color online) A layer of the Cmca structure
      at 300 GPa. The layers are arranged in an ABA fashion and the
      primitive unit cell contains 8 atoms. }
\end{figure}

This section reports the ground-state phase diagram of solid molecular
hydrogen in the pressure range corresponding to phases I, II and III
($100<P<500$ GPa).  Three different XC functionals were used to
calculate the enthalpies of the Cmca, C2/c, Cmca-12, and P$6_3$/m
structures found by recent PBE calculations\cite{Pickard} to be stable
(have the lowest enthalpy) at different pressures within this range.
These structures are illustrated in Figs.~\ref{P63m300}, \ref{C2c300},
\ref{Cmca12300}, and \ref{Cmca300}.

\begin{figure}[ht]
    \centering
    \includegraphics[width=0.45\textwidth]{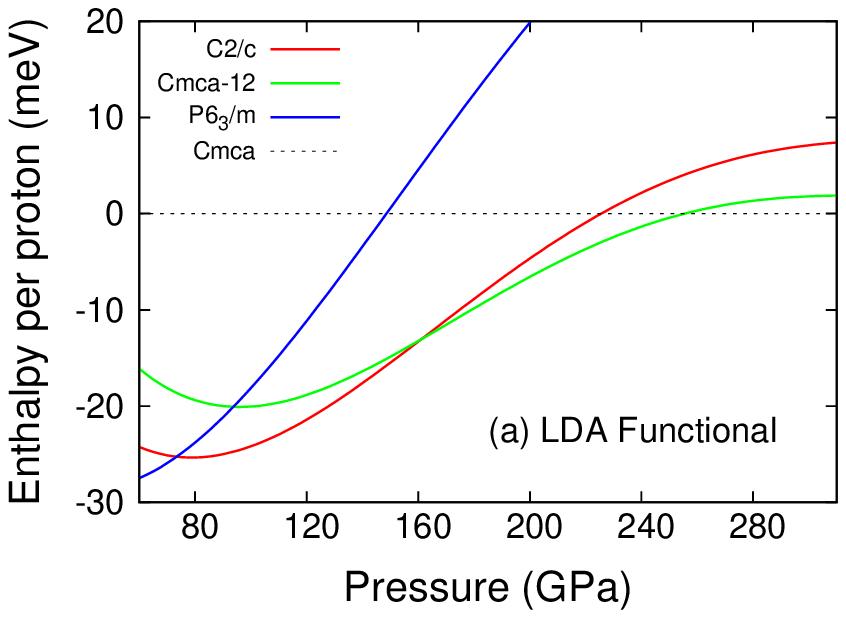} \\
    \includegraphics[width=0.45\textwidth]{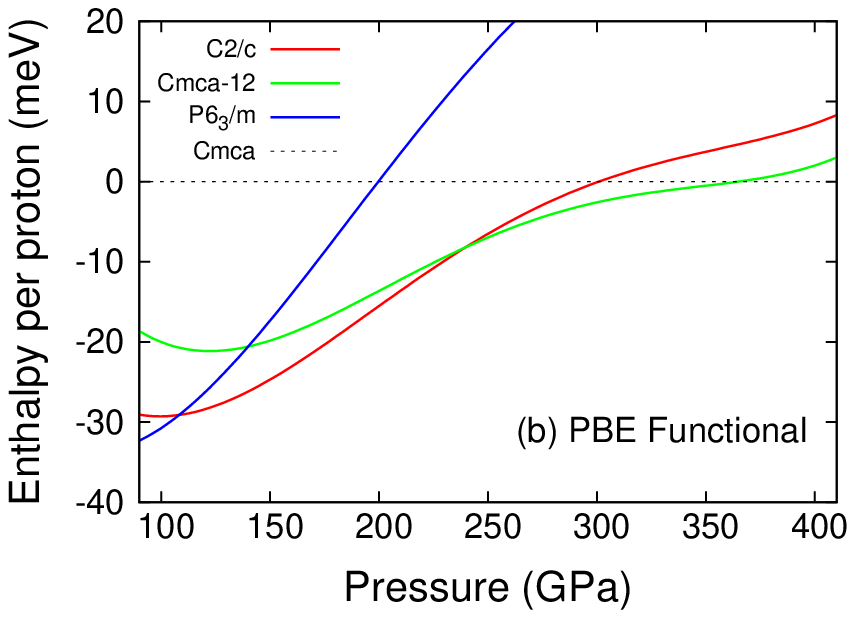} \\
    \includegraphics[width=0.45\textwidth]{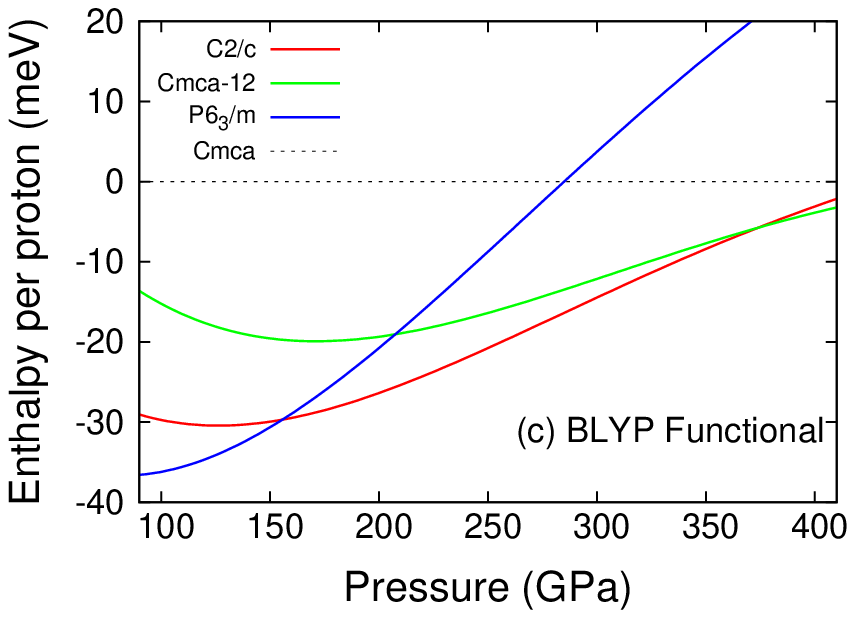}
    \caption{\label{Enthalpies}(color online). Enthalpy per proton as a
      function of pressure calculated using three different XC
      functionals: (a) LDA, (b) PBE, (c) BLYP. The static lattice
      enthalpies of three different insulating crystal structures are
      reported relative to the enthalpy of the metallic Cmca structure.
      The Cmca-12 structure, which is insulating at low pressure,
      transforms into the metallic Cmca structure at 260, 370 or 430 GPa
      according to the LDA, PBE and BLYP approximations, respectively.}
\end{figure}

Figure~\ref{Enthalpies} shows the static lattice enthalpy as a function
of pressure calculated using the LDA, PBE and BLYP functionals.
According to our LDA calculations, the P$6_3$/m, C2c, Cmca-12, and Cmca
phases are stable in the pressure ranges $<$ 70 GPa, 70--165 GPa,
165--260 GPa, and $>$ 260 GPa, respectively. Our PBE phase diagram is in
agreement with previous work,\cite{Pickard} indicating that the
P$6_3$/m, C2/c, Cmca-12, and Cmca phases are stable in the pressure
ranges $<$ 110 GPa, 110--245 GPa, 245--370 GPa, and $>$ 370 GPa, respectively.
According to experiment, the phase transition from phase I to phase II happens around
110 GPa and is characterized by a change in the low-frequency region of
the Raman and IR spectra.\cite{revmod} It is interesting that the
structural transition from P$6_3$/m to C2/c is also observed at 110 GPa
when the PBE functional is used in the static approximation.  The static
PBE phase diagram does not provide any information about the transition
from phase II to III, which is observed experimentally at around 150 GPa
and is accompanied by a large low-temperature discontinuity in the Raman
scattering and a strong rise in the IR molecular vibrons.\cite{revmod}
The static BLYP phase diagram, by contrast, identifies the transition
from phase II to III with the structural change from P$6_3$/m to C2/c.
As Fig.~\ref{Enthalpies}(c) illustrates, BLYP calculations predict that
the P$6_3$/m, C2c, Cmca-12, and Cmca phases are stable in the pressure
ranges of $<$ 160 GPa, 160--370 GPa, 370--430 GPa, and $>$ 430 GPa,
respectively.

\begin{figure}[ht]
    \centering
    \includegraphics[width=0.45\textwidth]{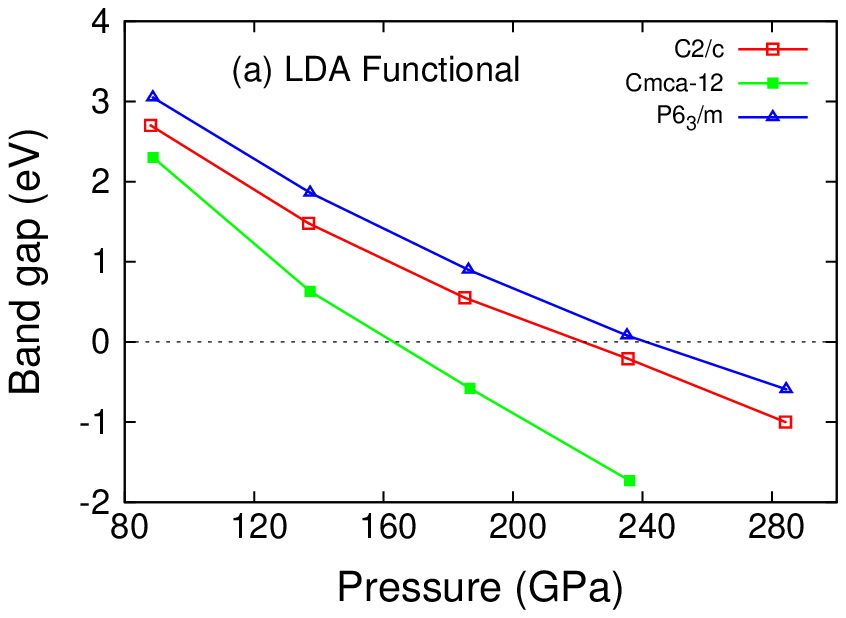} \\
    \includegraphics[width=0.45\textwidth]{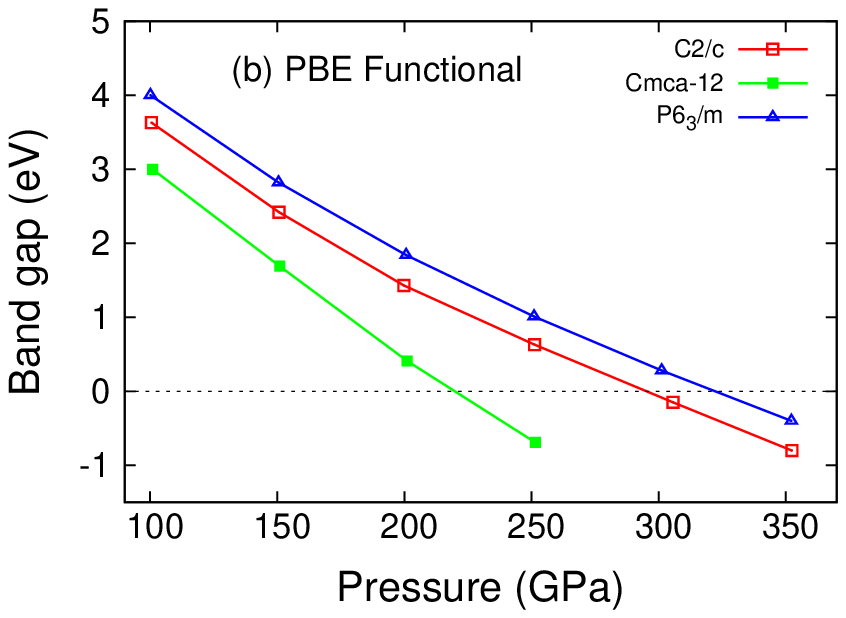} \\
    \includegraphics[width=0.45\textwidth]{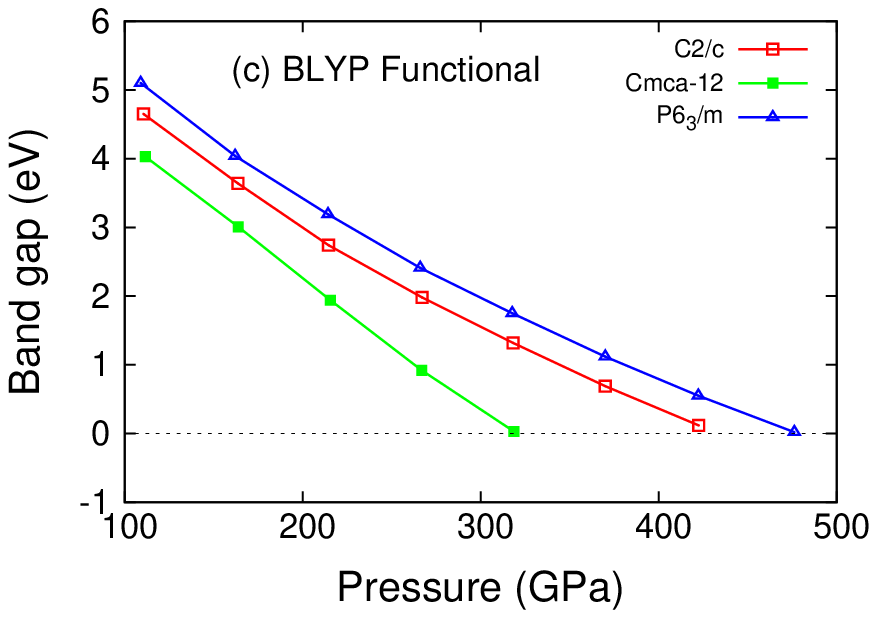}
    \caption{\label{XCgaps} (color online). Single-particle band gaps
      versus pressure calculated using (a) the LDA, (b) the PBE, and (c)
      the BLYP functionals. The BLYP band gap is larger than the LDA and
      PBE gaps and consequently yields larger metallization
      pressures. The P$6_3$/m and Cmca-12 phases have the highest and
      lowest metallization pressures, respectively, regardless of the XC
      functional used.}
\end{figure}

Figure~\ref{XCgaps} illustrates how the DFT band gaps of the Cmca-12,
C2/c and P$6_3$/m structures close as the pressure increases. The
results obtained using different XC functionals differ markedly. The DFT
band gap does not correspond to the measured quasi-particle band gap and
is usually much smaller, so these results are not expected to agree with
experiment; the true pressures at which the band gaps of these
structures close are likely to be significantly higher than suggested by
Fig.~\ref{XCgaps}.

Exact-exchange (EXX) DFT calculations yield band gaps 1--2 eV higher
than LDA gaps.\cite{Martin} Although this 1--2 eV difference was
obtained by comparing LDA and EXX calculations, Pickard\cite{Pickard}
investigated the effect of adding 1--2 eV to the PBE band gap of the
C2/c phase. Increasing the PBE gap by 1 eV or 2 eV and extrapolating to
the pressure at which the increased gap vanished gave metallization
pressures of 350 and 410 GPa, respectively.  Table \ref{dftgap} reports
the pressures of band-gap closure obtained using the LDA, PBE and BLYP
functionals with and without band-gap corrections of 1 eV, 1.5 eV and 2
eV. Use of the BLYP functional yields the highest calculated
metallization pressures for all three structures, even higher than the
pressures obtained by applying a 2 eV correction to the LDA gap.

The enthalpy-pressure relationships plotted in Fig.~\ref{Enthalpies}
show that the LDA favors the metallic Cmca phase relative to the
insulating phases, and hence that the structural transition to the Cmca
phase occurs at a lower pressure within the LDA.  Furthermore, Fig.\
\ref{XCgaps} shows that the band gaps of the insulating phases close at
lower pressures when the LDA functional is used. The BLYP functional, by
contrast, makes insulating structures relatively more stable than the
metallic Cmca structure and predicts that the insulating Cmca-12 phase
does not transform into the metallic Cmca phase until the pressure
reaches 430 GPa. All three XC functionals suggest that the P$6_3$/m
phase has the largest band gap and, consequently, the highest
metallization pressure of the three insulating structures
considered. The band gap of the Cmca-12 phase closes at the lowest
pressure.

\begin{table}
  \caption{\label{dftgap} 
    Calculated pressure in GPa at which the band gap of solid 
    molecular hydrogen in the Cmca-12, C2/c and P$6_3$/m structures 
    closes. The metallization pressure is obtained by plotting the 
    band gap as a function of pressure and extrapolating to the 
    point at which the gap vanishes. Band gaps obtained using the 
    LDA, PBE and BLYP XC functionals are shown. The effects of 
    adding band-gap corrections of 1.0, 1.5 and 2.0 eV to the LDA 
    results are also indicated.
  }
\begin{center}
\begin{tabular}{|c|c|c|c|c|c|c|}
\hline\hline
 Space Group & LDA & LDA$+1$ & LDA$+1.5$ & LDA$+2$ & PBE & BLYP\\
\hline
 Cmca-12  & 160 & 205 & 228 & 248 & 230 & 320\\
 C2/c     & 230 & 283 & 313 & 343 & 298 & 420\\
 P$6_3$/m & 240 & 305 & 336 & 366 & 325 & 480\\
\hline\hline
\end{tabular}
\end{center}
\end{table}

All three XC functionals also indicate that the insulating wide-band-gap
hexagonal structure P$6_3$/m is stable in the low-pressure
regime. Indeed, all three predict the same sequence of structural
transitions, P$6_3$/m $\rightarrow$ C2/c $\rightarrow$ Cmca-12
$\rightarrow$ Cmca, although at different pressures. They also agree
that the band gap of the Cmca-12 structure (which is insulating at low
pressure) closes before the transition to the Cmca structure (which is
metallic at low pressure) takes place.  However, the quantitative
disagreements between results obtained using different XC functionals
are so large that it is barely possible to compare the calculated DFT
phase diagram with experiment.  Assuming that the I $\rightarrow$ II and
II $\rightarrow$ III phase transitions observed experimentally at around
110 and 150 GPa are related to structural transformations, different
functionals ascribe qualitatively different structural transitions to
each. The PBE phase diagram predicts a transition from P$6_3$/m to C2/c
at around 105 GPa, which could be the I $\rightarrow$ II transition. The
LDA and BLYP phase diagrams show no structural transformations near 110
GPa but two different transformations, P$6_3$/m $\rightarrow$ C2/c and
C2/c $\rightarrow$ Cmca-12, in the pressure range 150--160 GPa, possibly
related to the II $\rightarrow$ III transition.  None of the three XC
functionals is able to explain both phase transitions.

\section {Dynamic results and discussion}

The treatment of nuclear quantum effects (ZPE) within DFT is a
challenging problem, but is especially important in hydrogen due to the
small nuclear mass. Indeed, we show below that including the proton ZPE
has a large effect on the phase diagram. DFT is also used to study other
properties that depend on lattice vibrations, including Raman and IR
spectra and the electron-phonon interaction.\cite{McMahon2} As in the
static case, the question arises as to how these dynamical aspects of
the behavior of high-pressure solid hydrogen are affected by the choice
of XC functional. Answering this question is more important in the
dynamic than the static case, since the experimental techniques most
widely used to study high-pressure solid hydrogen are Raman and IR
spectroscopy, both of which are interpreted using phonon
calculations.\cite{baroni}

\begin{figure}[ht]
    \centering
    \includegraphics[width=0.45\textwidth]{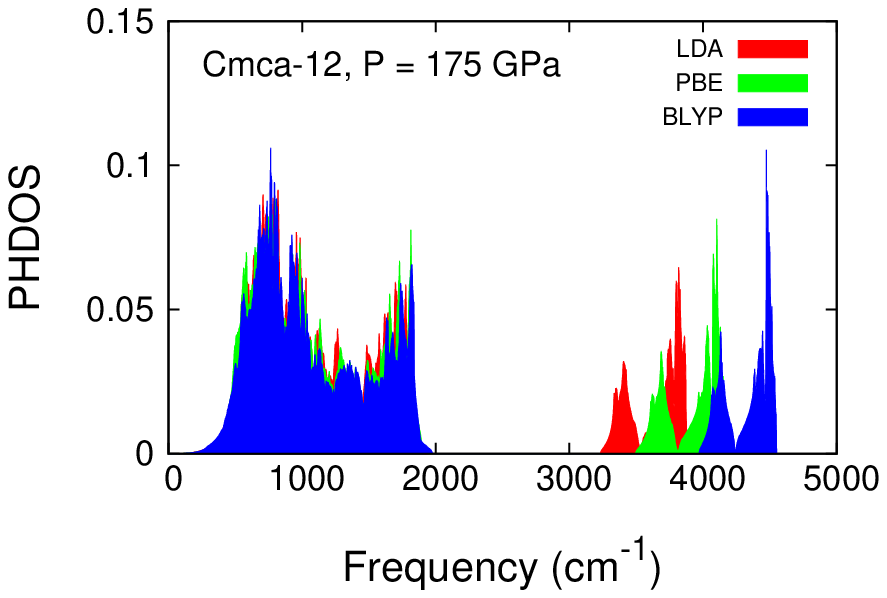} \\
    \includegraphics[width=0.45\textwidth]{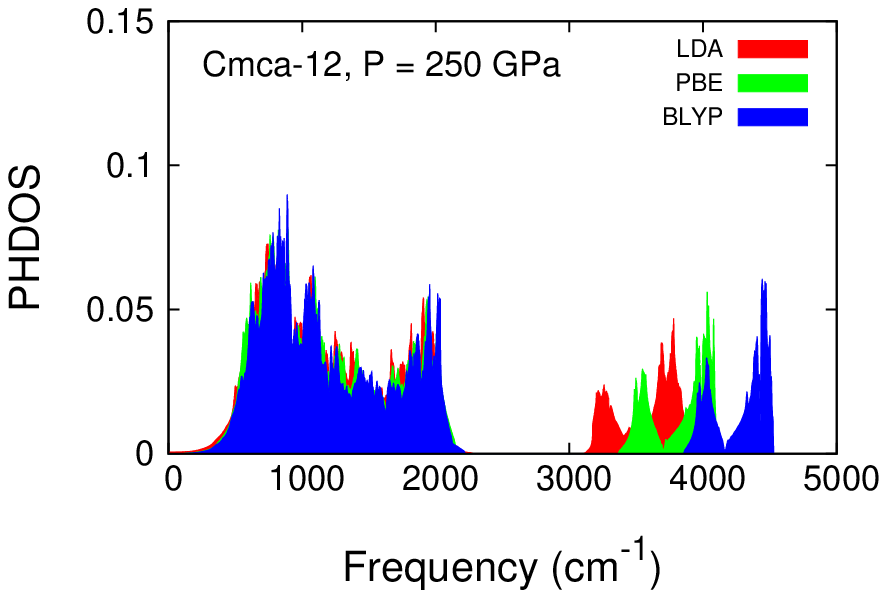} \\
    \includegraphics[width=0.45\textwidth]{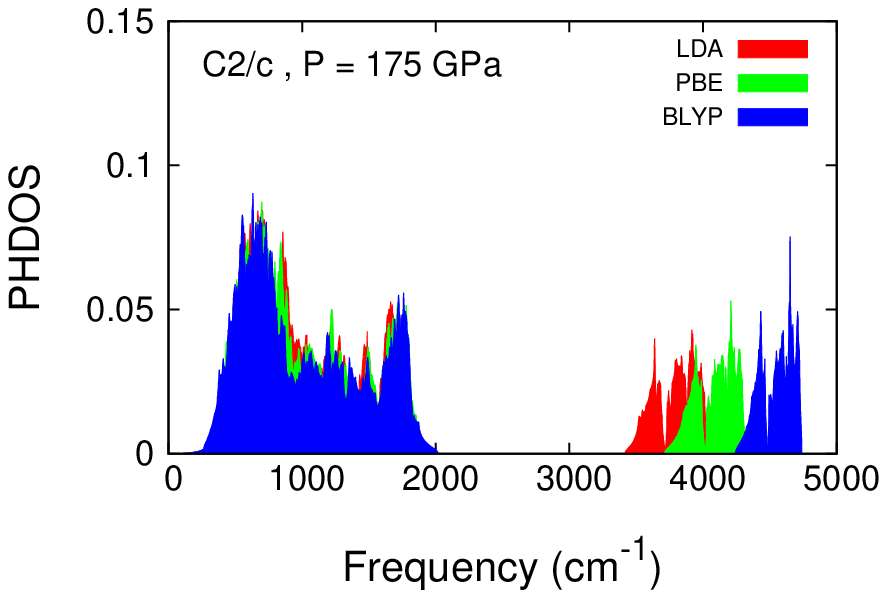} \\
    \includegraphics[width=0.45\textwidth]{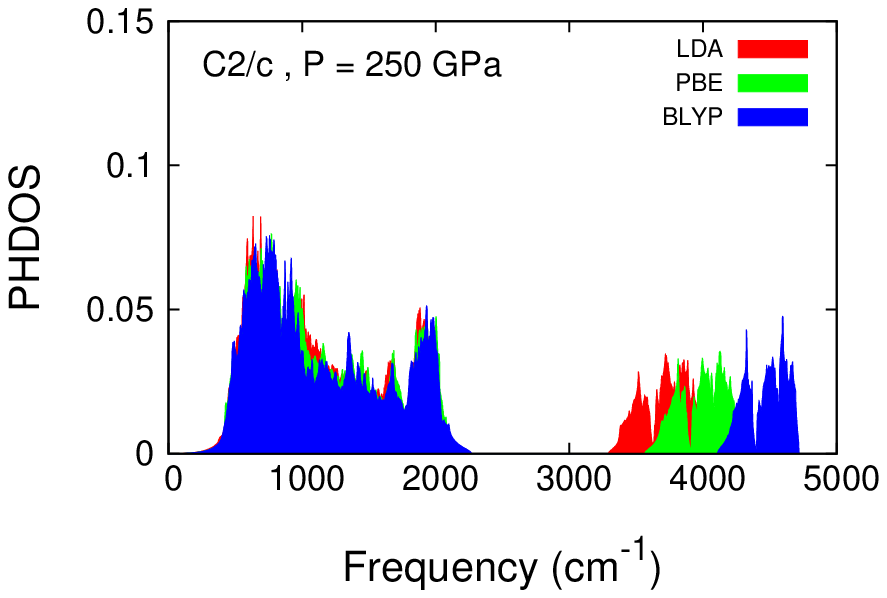}
    \caption{\label{phonon} (color online). The phonon densities of
      states of the Cmca-12 and C2/c structures calculated using the
      LDA, PBE and BLYP functionals at 175 and 250 GPa.  All three
      functionals yield almost the same low frequency ($<$ 2200
      cm$^{-1}$) phonon spectrum.  The frequencies of the vibron modes
      in the high-frequency regime ($>$ 3500 cm$^{-1}$) depend strongly
      on the choice of functional.}
\end{figure}

Figure~\ref{phonon} shows the phonon densities of states of the Cmca-12
and C2/c structures calculated using the LDA, PBE and BLYP XC
functionals. The principal effect of the choice of functional is on the
size of the phonon band gap, which depends on pressure and decreases as
the pressure increases.  As in the case of the electronic band gap, the
BLYP and LDA functionals give the largest and smallest phonon gap,
respectively.  Surprisingly, at low frequencies ($<$ 2200 cm$^{-1}$),
the phonon spectra obtained using the three different XC functionals are
quite similar.  The XC functional has a much larger effect in the
high-frequency regime ($>$ 3500 cm$^{-1}$). As Fig.~\ref{phonon} shows,
the LDA, PBE and BLYP functionals all predict two strong modes above the
phonon gap, but at significantly different frequencies. The
uncertainties in the frequencies of these modes complicate the
comparison of computational and experimental Raman and IR spectra. To
illustrate this difficulty, we have investigated the sensitivity of the
calculated IR spectrum of C2/c, Cmca-12, and P$6_3$/m phases at 175 GPa
to the choice of XC functional. 

\begin{figure}[ht]
    \centering
    \includegraphics[width=0.45\textwidth]{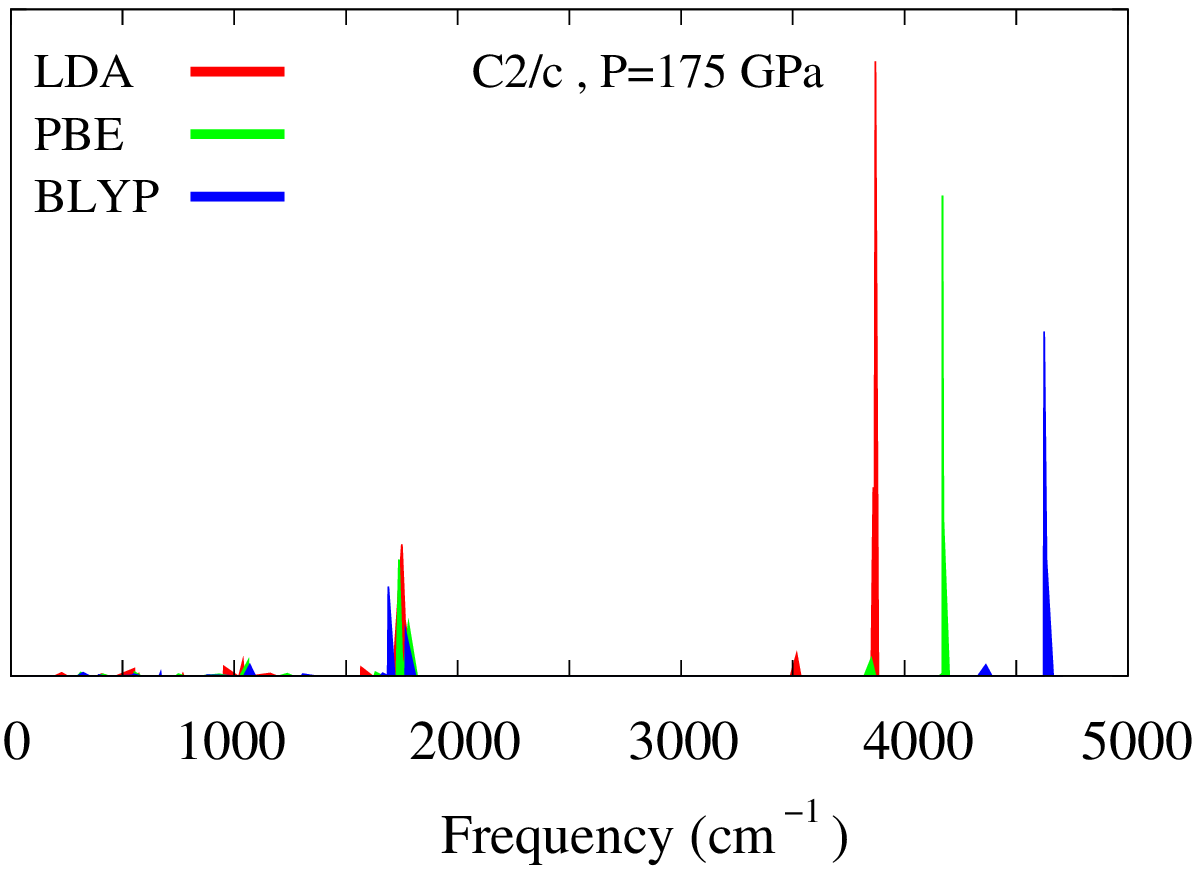} 
    \includegraphics[width=0.45\textwidth]{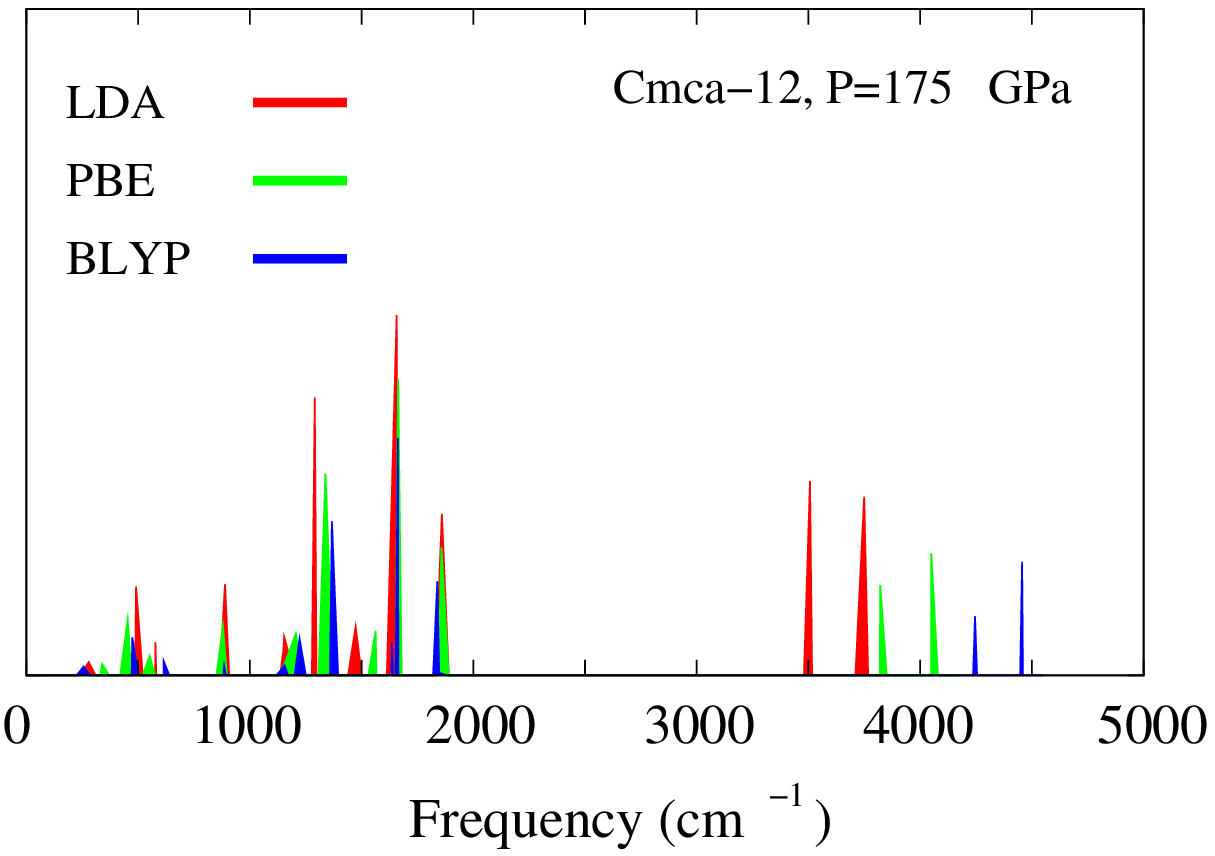} 
    \includegraphics[width=0.45\textwidth]{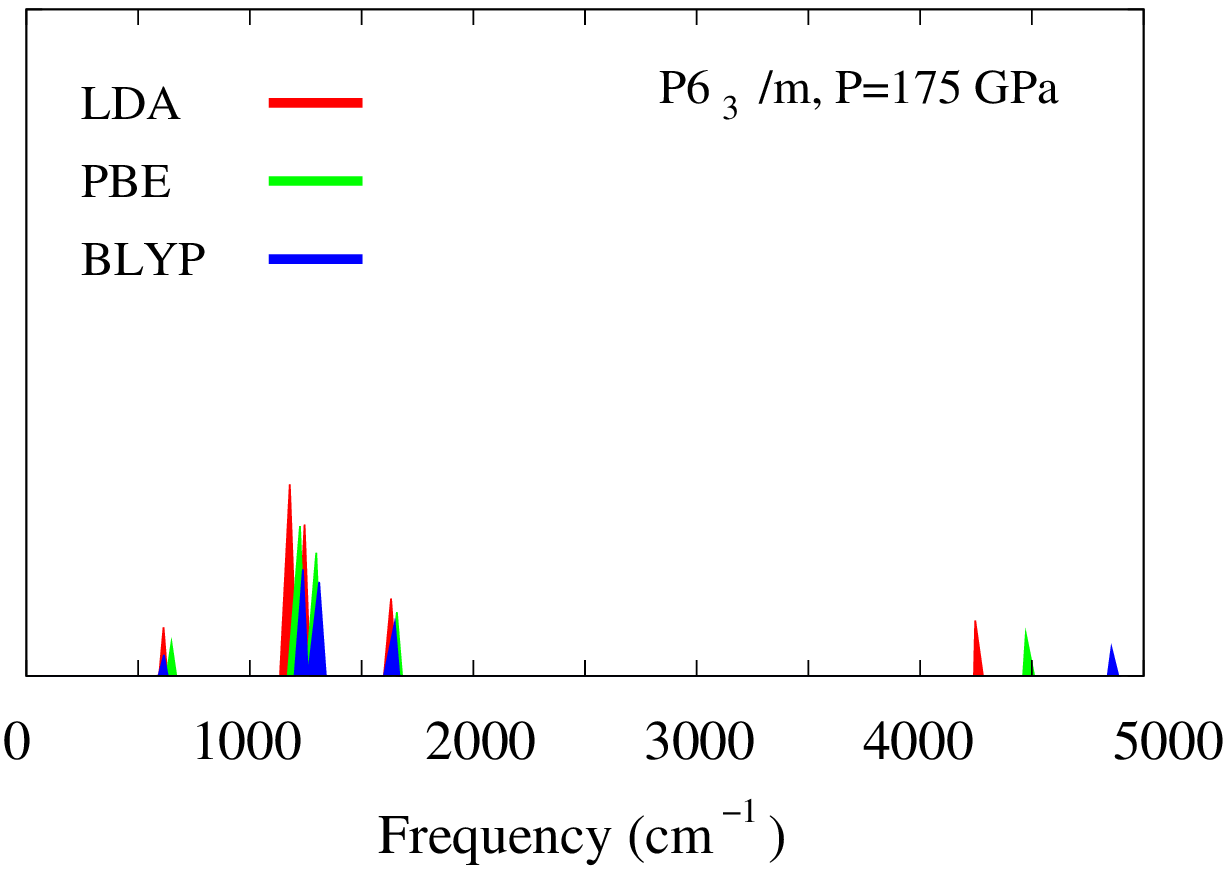} 
    \caption{\label{IRS} (color online). Simulated IR spectra of the
      C2/c, Cmca-12 and P$6_3$/m phases at 175 GPa. The intensity scale
      used for the P$6_3$/m phase is double that used for the Cmca-12
      and P$6_3$/m phases.}
\end{figure}

As illustrated in Fig.~\ref{IRS}, the computed IR spectrum of the C2/c
phase exhibits a single strong vibron peak, although the peak intensity
depends on the choice of XC functional. The peak frequencies calculated
using the LDA, PBE, and BLYP functionals are approximately 3800, 4200,
and 4600 cm$^{-1}$, respectively. In the case of the Cmca-12 phase, all
three functionals produce two peaks of similar intensity in the high
frequency regime. The LDA peak frequencies are 3500 and 3750 cm$^{-1}$;
the PBE frequencies are 3820 and 4050 cm$^{-1}$; and the BLYP
frequencies are 4250 and 4460 cm$^{-1}$.  The simulated IR spectrum of
the P$6_3$/m phase shows a single high-frequency peak, which appears at
4250, 4470, or 4860 cm$^{-1}$ for the LDA, PBE, and BLYP functionals,
respectively. 

Experimental IR spectra\cite{Hemley12} taken at 158 GPa show a single
strong vibron peak with a frequency of 4400 cm$^{-1}$. As the pressure
increases, the frequency of this peak decreases, reaching 4300 cm$^{-1}$
at 208 GPa, suggesting a frequency close to 4350 cm$^{-1}$ at P=175 GPa.
Experiment also shows\cite{Hemley12} that the vibron intensity below 150
GPa is approximately three orders of magnitude lower than at higher
pressures.  Although no XC functional yields the experimental IR
frequency accurately, all indicate that the C2/c phase has a very strong
IR peak in roughly the right frequency range.  As pointed out by Pickard
and Needs,\cite{Pickard} this observation is consistent with the
suggestion that phase III has the C2/c structure. Regardless of the
choice of XC functional, the C2/c vibron peak is much more intense than
the vibron peaks for the other structures studied.

The effect of ZPE on the phase diagram is normally studied by applying a
simple, additive, pressure-dependent correction to the static
ground-state total energy. The zero-point motion is assumed to be
harmonic and the phonon frequencies calculated as a function of volume
using DFT. This was the procedure adopted here. The limitations of the
harmonic approximation can be overcome using DFT-based path-integral
molecular dynamics,\cite{Li13} but this is very costly and still relies
on the accuracy of the XC functional, which is clearly in
doubt. Ground-state QMC simulations treating both the electrons and
protons as quantum mechanical particles\cite{Ceperley87} do not rely on
an approximate XC functional but are even more expensive. Furthermore,
only a few QMC simulations of this type have ever been attempted and
there are good reasons to doubt their accuracy in practice.

\begin{figure}[ht]
    \centering
    \includegraphics[width=0.45\textwidth]{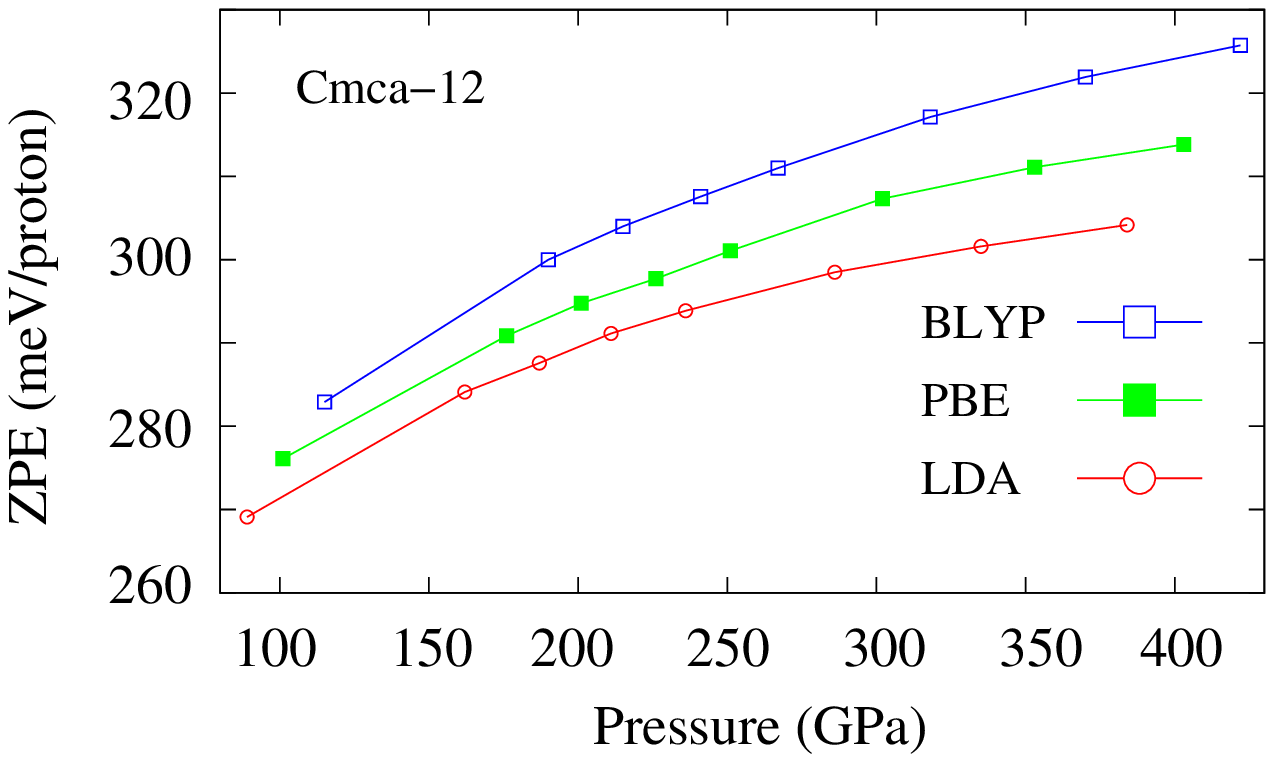} \\
    \includegraphics[width=0.45\textwidth]{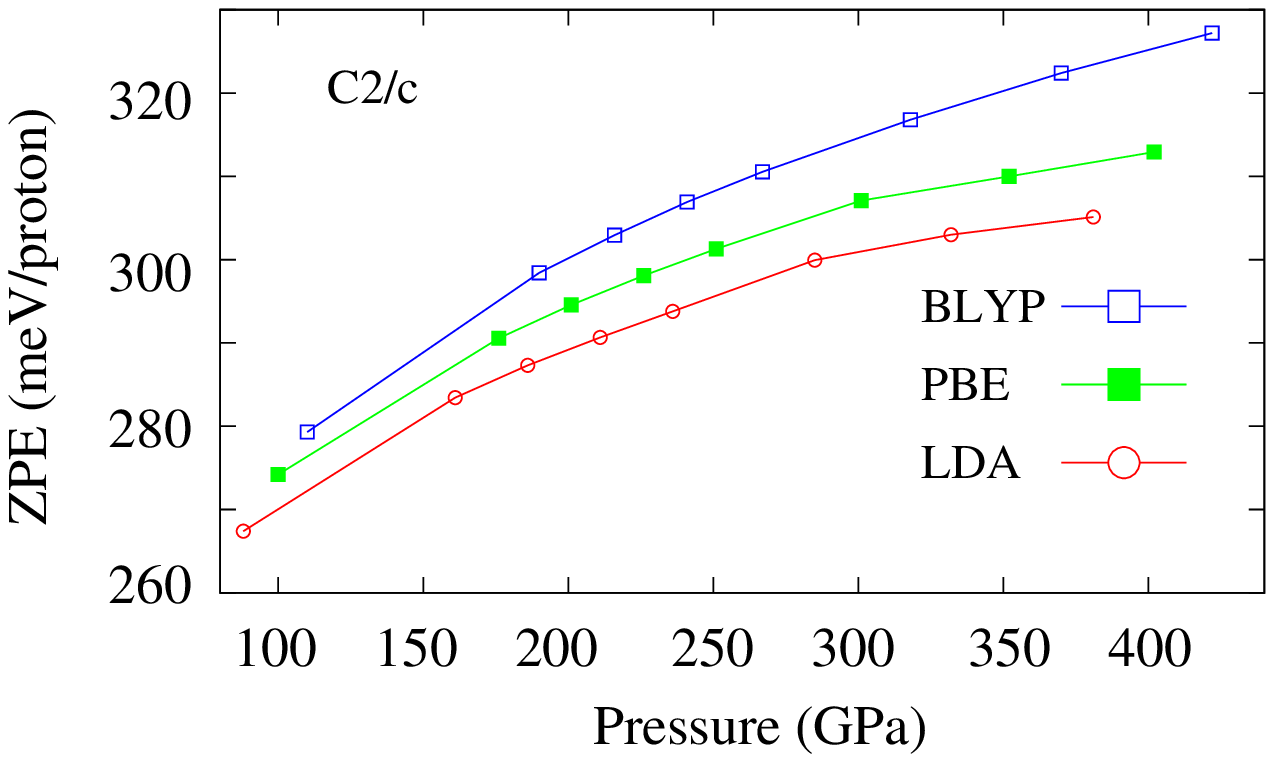} \\
    \includegraphics[width=0.45\textwidth]{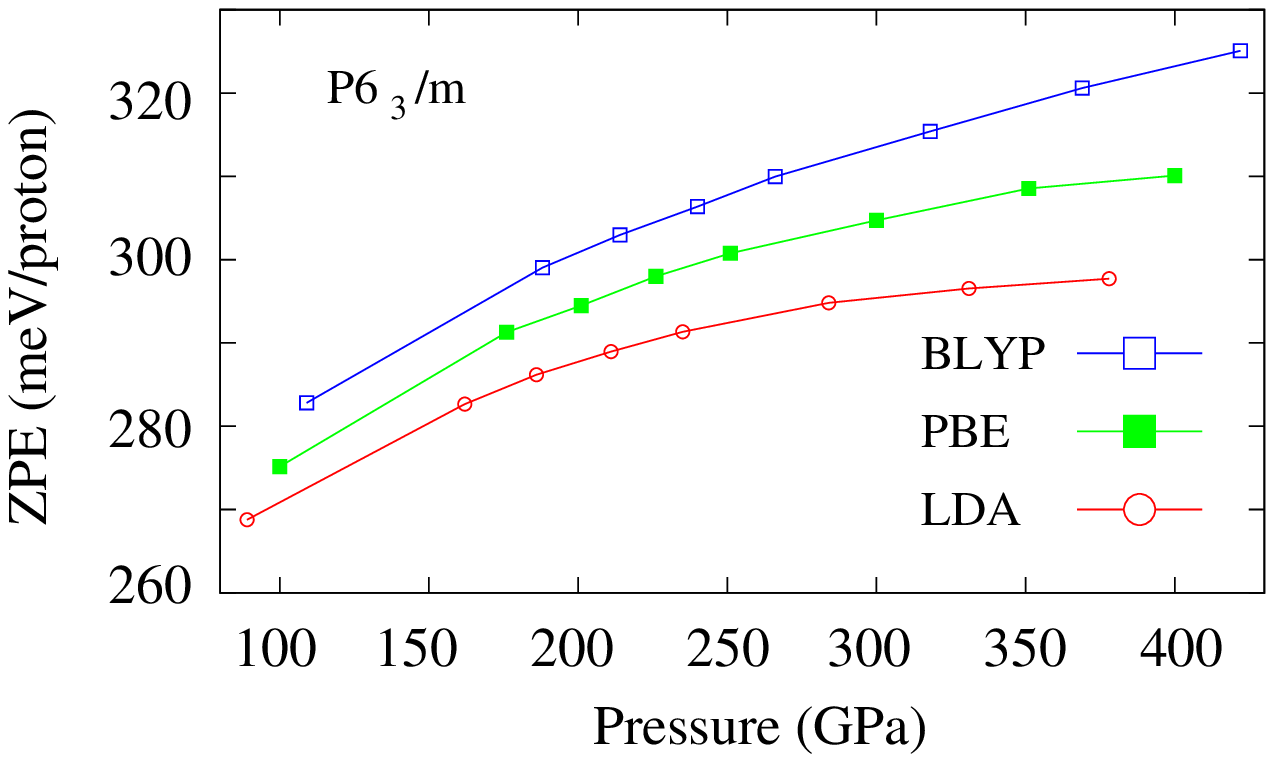} \\
    \includegraphics[width=0.45\textwidth]{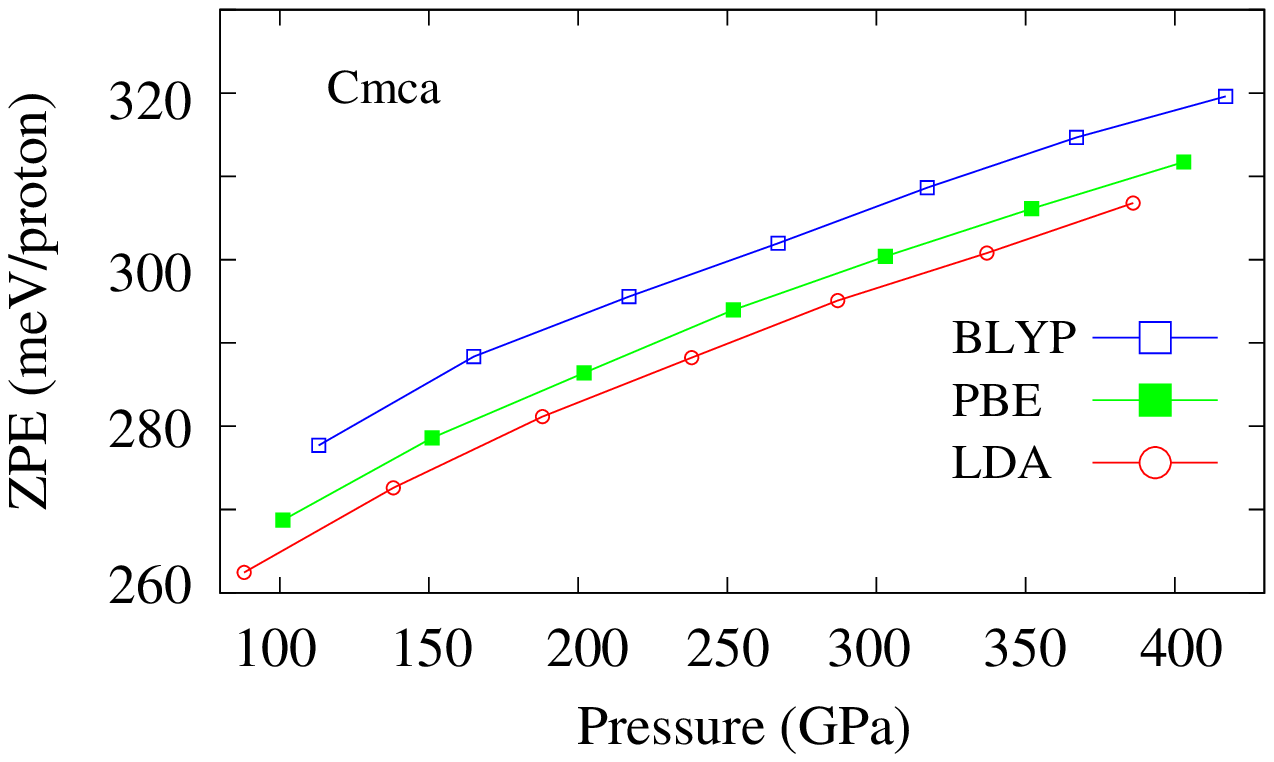} 
    \caption{\label{ZPEs} (color online). The proton zero-point energies
      of the Cmca-12, C2/c, P$6_3$/m and Cmca crystal structures
      within the pressure range of phases I,II, and III as calculated using the
      LDA, PBE and BLYP functionals. The ZPE per proton increases with
      pressure.  At any given pressure, the difference between ZPEs
      calculated using the BLYP and LDA functionals is more than 10
      meV/proton; this is large in comparison with the few meV/proton
      enthalpy differences between structures.}
\end{figure}

Figure~\ref{ZPEs} summarizes the results of our proton ZPE calculations
using the LDA, PBE and BLYP functionals. The proton ZPEs of all the
structures investigated, Cmca-12, C2/c, P$6_3$/m, and Cmca, are shown in
the pressure range of phases I, II, and III. It is clear that the ZPE
increases with increasing pressure regardless of the crystal structure
or choice of XC functional, and that the BLYP and LDA functionals give
the largest and smallest ZPEs in all cases.  For the insulating phases,
Cmca-12, C2/c, and P$6_3$/m, increasing the pressure also increases the
differences between the proton ZPEs obtained using different XC
functionals.  For the metallic Cmca phase, these differences are
approximately independent of pressure.  The differences between the
proton ZPEs of the candidate crystal structures can be comparable to the
differences between their static total energies. Therefore, structure
searching algorithms including an approximate ZPE should be considered.

The crystal structures investigated here were all mechanically stable
and fully relaxed in the supercells used for structure searching, but
the existence of small imaginary phonon frequencies indicates that some
may be weakly unstable with respect to structural distortions on longer
length scales.\cite{Pickard} The imaginary phonon frequencies are
relatively small, however, and their number can be reduced by
time-consuming small relaxations of the structures in larger supercells.

\begin{figure}[ht]
    \centering
    \includegraphics[width=0.45\textwidth]{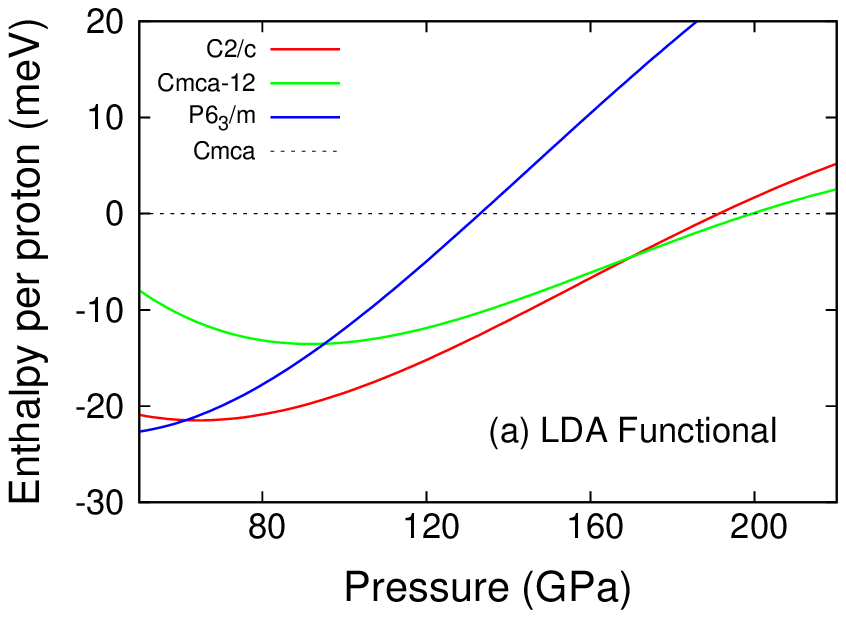} \\
    \includegraphics[width=0.45\textwidth]{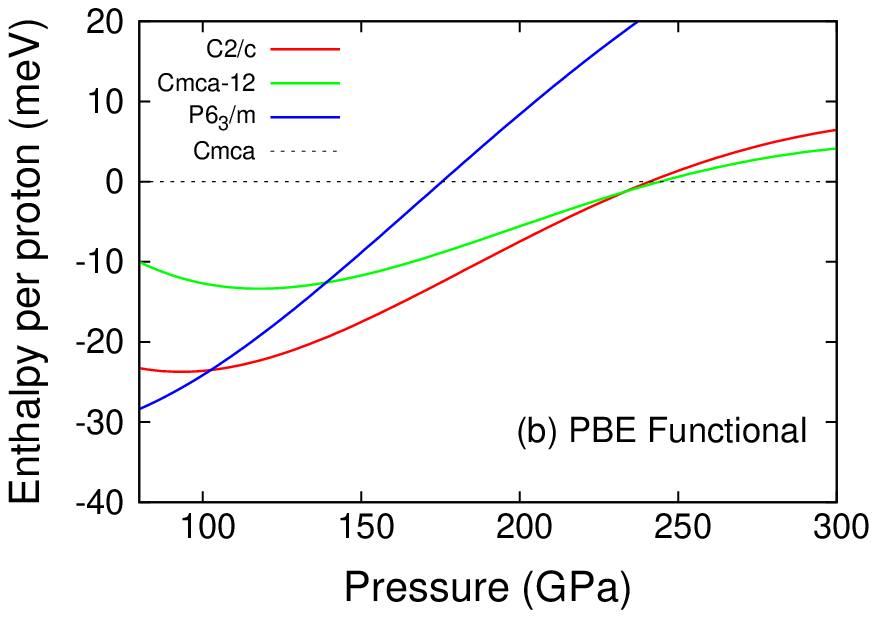} \\
    \includegraphics[width=0.45\textwidth]{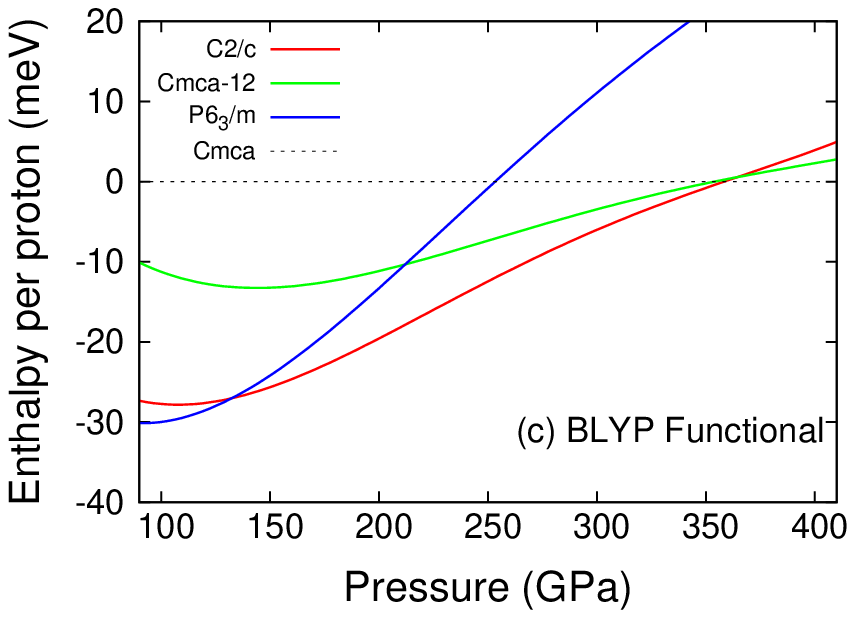}
    \caption{\label{EnthalpiesZPE}(color online). Enthalpy per proton
     included ZPE (dynamic enthalpy) as a
      function of pressure calculated using three different XC
      functionals: (a) LDA, (b) PBE, (c) BLYP. The dynamic lattice
      enthalpies of three different insulating crystal structures are
      reported relative to the enthalpy of the metallic Cmca structure.}
\end{figure}

The static phase diagrams presented in Fig.~\ref{Enthalpies} assume that
the protons are infinitely massive. By adding the proton ZPE to the
static results we obtain the dynamic phase diagrams reported in
Fig.~\ref{EnthalpiesZPE}.  The LDA dynamic phase diagram shows three
phase transformations: P$6_3$/m to C2/c at around 50 GPa, C2/c to
Cmca-12 at around 170 GPa, and the Cmca-12 to Cmca metal-insulator
transition at around 200 GPa. The PBE and BLYP dynamic phase diagrams
place the P$6_3$/m to C2/c transition at 110--130 GPa.  This suggests
that the I-II phase transition observed experimentally at 110 GPa
corresponds to the P$6_3$/m to C2/c structural transformation, even
though the IR data provides suggests that C2/c corresponds to phase III.
The dynamic PBE and BLYP phase diagrams agree that the Cmca-12 structure
is never stable in the pressure range associated experimentally with
phase III and that the C2/c structure transforms directly, or almost
directly, into the metallic Cmca structure. The transition to Cmca-12
proposed by Pickard and Needs\cite{Pickard} has not been observed
experimentally\cite{Hemley12} and a recent metadynamics calculation
\cite{Liu} also suggests that, at finite temperature, phase III may
transform to Cmca without passing through Cmca-12. If the Cmca-12 phase
is ever stable, it is only within a very narrow window just below the
pressure of the transition to the metallic Cmca structure.

The inclusion of the effects of proton ZPE substantially decreases the
metallization pressure. Although the PBE and BLYP dynamic phase diagrams
can both be interpreted as in agreement with experimental results
concerning metallization, they do not agree with the \emph{same}
experimental results. Eremets and Troyan claim that \cite{Eremets}
hydrogen transforms into a metal at 260--270 GPa: the conductance
increases sharply and changes little on increasing the pressure to
300~GPa or cooling to 30~K; and the sample reflects light well. As is
shown in Fig.~\ref{EnthalpiesZPE}(b), the PBE dynamic phase diagram
supports this claim: C2/c remains the most stable structure throughout
phases II and III and transforms directly or almost directly into the
metallic Cmca phase at 240--250 GPa.  On the other hand, Hemley and
co-workers\cite{Hemley12} see no evidence of a metallic state up to 360
GPa, but report electronic properties consistent with semi-metallic
behaviour.  The BLYP dynamic phase diagram, Fig.~\ref{EnthalpiesZPE}(c),
shows a metal-insulator transition at 350--360 GPa, related to a phase
transition from the C2/c structure to the Cmca structure. The dynamic
phase diagrams also suggest that metallization of high-pressure solid
hydrogen happens through a molecular structural transformation, not
band-gap closure. The pressure at which the band gap of the C2/c phase
closes, as shown in Table~\ref{dftgap} and also as calculated using the
more accurate $GW$ approximation,\cite{me2} is considerably higher than
the pressure at which the transition to the metallic Cmca structure
takes place. Since metallization is likely to take place via a
structural transformation, the failure of DFT to produce accurate band
gaps is unlikely to be relevant.

Last, but not least, we ask why DFT is unable to reproduce the phase
diagram of high-pressure solid hydrogen.  We believe that there are two
main problems. One is caused by the very small enthalpy differences of a
few meV per proton between phases. Exhaustive tests for molecules, where
accurate results can be obtained using quantum chemical methods, have
shown that no current XC functional is consistently capable of achieving
such accuracy.\cite{cohen} The second is the many-body self-interaction
(XC-SI) error present in the XC functionals used here.  A systematic
investigation\cite{polo} found the XC-SI errors of the LDA, GGA, and
BLYP total energies of a single H$_2$ molecule to be 1.264, -0.126, and
0.0846 eV, respectively. These values are more than two orders of
magnitude larger than the DFT enthalpy differences between the crystal
structures of high-pressure solid hydrogen. The GGA functional has a
lower X-SIE than the LDA functional, but the C-SIE, which is negative by
definition, is larger than the X-SIE in this case. The XC-SIE of the
BLYP functional is relatively small because the LYP functional by
construction does not suffer from the C-SIE. In previous work\cite{me1}
we showed that the use of hybrid XC functionals, which generally suffer
from smaller XC-SI errors than pure density functionals,\cite{cohen}
changes the phase diagram considerably.

\section {Conclusion}

This paper reported DFT calculations using the LDA, PBE and BLYP XC
functionals of various proposed structures of solid molecular hydrogen
at low temperature and high pressure. Static results obtained using all
three XC functionals are consistent with the following scenario: at the
highest pressures considered here, the Cmca-12 structure, which is
insulating at low pressure but becomes semi-metallic with increasing
pressure, transforms into the Cmca structure, which is metallic even at
low pressure. Since the Cmca-12 and Cmca structures have the same space
group, differing only in the number of hydrogen molecules per unit cell,
this seems reasonable. More generally, as the pressure is increased, the
band gaps of the insulating structures of solid hydrogen become
smaller. As the band gaps decrease, we find that structures with wider
gaps become progressively less stable relative to structures with
smaller gaps, and a series of transitions to structures with smaller
gaps takes place. This process culminates in the transition to the
metallic Cmca structure.

The above scenario changes significantly when the effects of proton ZPE
are added to the static phase diagram.  Our ZPE calculations were based
on the harmonic approximation, which is standard but may be problematic.
However, it is clear from our results that the effects of proton ZPE are
essential and must be taken into account, even if only approximately.
Surprisingly, but perhaps fortuitously, the agreement between the PBE
and BLYP dynamic phase diagrams is much better than the agreement
between the corresponding static phase diagrams. In particular, both
dynamic phase diagrams suggest that the experimentally observed I-II
phase transition corresponds to a structural transformation from
P$6_3$/m to C2/c, that C2/c is the most stable of the structures
considered over the whole pressure range of phases II and III, and that
metallization happens via a molecular structural transformation from
C2/c to Cmca without passing through Cmca-12. The main difference
between the PBE and BLYP dynamic phase diagrams is the pressure of the
C2/c to Cmca transition, which the PBE functional places at 240--250 GPa
and the BLYP functional places at 350--360 GPa. As discussed above, both
results can be interpreted as being in agreement with recent
experimental results, but different and conflicting ones. The
low-frequency ($<$ 2200 cm$^{-1}$) parts of the phonon spectra of all
structures considered are surprisingly independent of the choice of XC
functional, but this is not true of the high-frequency vibronic
parts. In general, it appears that the BLYP functional produces stiffer
phonons than the PBE functional and considerably stiffer phonons than
the LDA functional.

This work focused on four specific crystal structures recently proposed
on the basis of PBE calculations.\cite{Pickard} Although the phase
diagrams obtained using different XC functionals differed substantially
in quantitative terms, especially at the static level, the four
structures always appeared in the same order as the pressure was
increased. We think it likely, however, that if we had considered a
wider range of structures, we would have seen qualitative as well as
quantitative differences between the phase diagrams calculated using
different XC functionals. DFT is probably not accurate enough to allow
reliable comparisons of the stabilities of the insulating,
semi-metallic, and metallic phases of solid molecular hydrogen. Indeed,
since simple DFT calculations fail to describe van der Waals
interactions and severely underestimate band gaps, it would be
unreasonably optimistic to expect such comparisons to be reliable. We
conclude that using DFT results as the basis of our understanding of
high-pressure solid hydrogen is likely to lead to serious errors. To
obtain reliable results, it is almost certainly necessary to go beyond
DFT and take many-body effects properly into account.\cite{me2} In any
case, it is clear that more work needs to be done, both theoretically
and experimentally, to understand solid molecular hydrogen at low
temperature.


\end{document}